\documentclass[aps,prb,twocolumn,floatfix,amssymb,superscriptaddress]{revtex4}

\usepackage{amsmath}
\usepackage{graphicx}

\newcommand{\bk}{\mathbf{k}}
\newcommand{\half}{\mbox{$\frac{1}{2}$}}
\newcommand{\pdag}{\phantom{\dag}}

\begin{document}

\title{Signatures of quantum phase transitions in parallel quantum dots:
Crossover from local-moment to underscreened spin-1 Kondo physics}

\date{\today}

\author{Arturo Wong}
\affiliation{Department of Physics and Astronomy, Nanoscale and Quantum
Phenomena Institute, Ohio University, Athens, Ohio 45701, USA}
\affiliation{Department of Physics, University of Florida, P.O.\ Box 118440,
Gainesville, Florida 32611, USA}
\author{W.\ Brian Lane}
\affiliation{Department of Physics, University of Florida, P.O.\ Box 118440,
Gainesville, Florida 32611, USA}
\affiliation{Department of Physics, Jacksonville University, 2800 University
Boulevard North, Jacksonville, Florida 32211, USA}
\author{Luis G.\ G.\ V.\ Dias~da~Silva}
\affiliation{Instituto de F\'{i}sica, Universidade de S\~{a}o Paulo,
C.P.\ 66318, 05315-970 S\~{a}o Paulo, SP, Brazil}
\author{Kevin Ingersent}
\affiliation{Department of Physics, University of Florida, P.O.\ Box 118440,
Gainesville, Florida 32611, USA}
\author{Nancy Sandler}
\author{Sergio E.\ Ulloa}
\affiliation{Department of Physics and Astronomy, Nanoscale and Quantum
Phenomena Institute, Ohio University, Athens, Ohio 45701, USA}

\begin{abstract}
We study a strongly interacting ``quantum dot 1'' and a weakly interacting
``dot 2'' connected in parallel to metallic leads. Gate voltages can drive the
system between Kondo-quenched and non-Kondo free-moment phases separated by
Kosterlitz-Thouless quantum phase transitions. Away from the immediate vicinity
of the quantum phase transitions, the physical properties retain signatures
of first-order transitions found previously to arise when dot 2 is strictly
noninteracting. As interactions in dot 2 become stronger relative to the
dot-lead coupling, the free moment in the non-Kondo phase evolves smoothly from
an isolated spin-one-half in dot 1 to a many-body doublet arising from the
incomplete Kondo compensation by the leads of a combined dot spin-one. These
limits, which feature very different spin correlations between dot and lead
electrons, can be distinguished by weak-bias conductance measurements performed
at finite temperatures.
\end{abstract}


\maketitle

\section{Introduction}

Semiconductor quantum dots afford a level of experimental control that has made
them the premier setting\cite{Kouwenhoven} in which to investigate the Kondo
effect, \textit{i.e.}, the many-body screening of a local moment by delocalized
electrons. In recent years, interest has turned from Kondo physics in single
dots to similar phenomena in more complex structures such as double-dot
devices,\cite{DQD-expts,Potok:07} where quantum phase transitions (QPTs) have
been predicted\cite{Hofstetter:01,DQD-QPT-theory,Zitko:06,Dias:06,Dias:08,%
Dias:09,Logan:09} and possibly observed.\cite{Potok:07}

Kondo physics in two spin-degenerate quantum dots (or two levels within a
single dot) connected in parallel to the same single-channel leads has been
investigated from a number of perspectives. The combined spin of the two
localized levels can be tuned between singlet and triplet configurations by
adjusting a magnetic field\cite{Tarucha:00} or gate voltages.\cite{Kogan:03}
When coupled to leads, such setups exhibit enhanced conductance near the
singlet-triplet level crossing,\cite{vanderWiel:02,Kogan:03,ST-theory} with QPTs
of the Kosterlitz-Thouless type.\cite{Hofstetter:01,Vojta:02} Another theme that
has received considerable attention is the role of interference between
different current paths in modulating the conductance through parallel
quantum-dot setups\cite{DQD-interference,Dias:06,Dias:08} or pairs of dots
embedded in the arms of an Aharanov-Bohm ring.\cite{DQD-AB-interference,Dias:09}

Theoretical studies of parallel double quantum dots have overwhelmingly focused
on the limit in which each dot has strong Coulomb interactions and can acquire a
magnetic moment. Such systems exhibit two phases\cite{Zitko:06,Logan:09}: a
Fermi-liquid phase with a singlet ground state, and a ``singular Fermi liquid''
phase having a residual spin-\half\ arising from an underscreened spin-1 Kondo
effect.\cite{USC} These phases are separated by lines of Kosterlitz-Thouless
QPTs broken by first-order QPTs at points of exact equivalence between the
dots.\cite{Logan:09}

Parallel doublet dots in a very different limit, where ``dot 1'' has strong
interactions but ``dot 2'' is strictly noninteracting (and hence nonmagnetic),
have been shown\cite{Dias:06} to realize the pseudogap Kondo
effect,\cite{pseudogap,Ingersent-pseudo} in which a magnetic impurity couples
to a conduction band having a density of states that vanishes in power-law
fashion at the Fermi energy. This reduction of the low-energy density of
states inhibits the Kondo effect unless the effective impurity-band exchange
coupling exceeds a critical value. The Kondo-screened phase is separated from
a non-Kondo local-moment phase by first-order QPTs that exhibit clear
signatures in finite-temperature transport.\cite{Dias:08}

In this work we explore the connection between limits described in the
previous two paragraphs by considering the effect of increasing the dot-2
Coulomb interaction $U_2$ from zero. A free-moment phase with an unquenched
spin-\half\ occupies a region of parameter space that grows with $U_2$ and is
separated from a surrounding strong-coupling phase by Kosterlitz-Thouless QPTs.
For $U_2\lesssim\Gamma_2$---the level width of dot 2 due to its coupling to the
leads---the properties retain signatures of the $U_2=0$ pseudogap Kondo
physics, while for $U_2\gg\Gamma_2$ there is a smooth crossover to
the heavily studied limit of two strongly interacting dots. These two regimes,
both exhibiting singular Fermi liquid behavior with very different dot-lead
entanglements, can be distinguished through weak-bias conductance
measurements at experimentally accessible temperatures. In experiments, it is
impractical to adjust $U_2$ by orders of magnitude, but the crossover from
$U_2\ll\Gamma_2$ to $U_2\gg\Gamma_2$ can be accessed by tuning $\Gamma_2$ via
gate voltages. The setup therefore has great potential for investigation of
QPTs and of entanglement in singular Fermi liquids, which lie on the borderline
between regular Fermi liquids and non-Fermi liquids.\cite{Mehta:05}

The double-quantum-dot setup and its phase diagram are described in
Sec.\ \ref{sec:model}. Section \ref{sec:zero_vs_weak_U2} compares the cases
$U_2=0$ and $U_2=\Gamma_2$, the latter typifying the behavior for a weakly
correlated dot 2, while Sec.\ \ref{sec:small_vs_large_U2} addresses the
crossover from weak to strong dot-2 interactions.
The results are summarized in Sec.\ \ref{sec:summary}. 

\section{Model and phase diagrams}
\label{sec:model}

\begin{figure}
\centerline{\includegraphics[width=0.8\columnwidth]{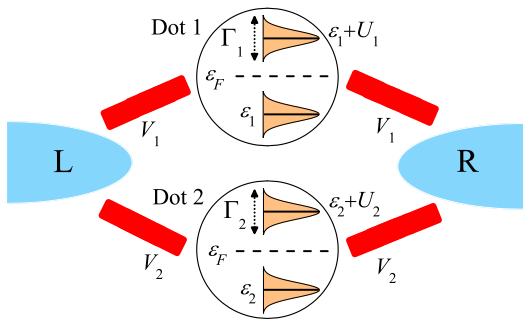}}
\caption{\label{Fig1} (Color online)
Schematic of the parallel double-quantum-dot setup considered in this work.}
\end{figure}

We consider an equilibrium system represented schematically in Fig.\ \ref{Fig1}
and modeled by a generalized Anderson Hamiltonian
\begin{equation}
\label{H_full}
H = H_{\text{leads}}+H_{\text{dots}}+H_{\text{mix}}.
\end{equation}
Here,
\begin{equation}
\label{H_leads}
H_{\text{leads}}=\sum_{j,\bk,\sigma} \epsilon_{j\bk}
  c_{j\bk\sigma}^{\dag} c_{j\bk\sigma}^{\pdag}
\end{equation}
represents the left ($L$) and right ($R$) leads, with $c_{j\bk\sigma}$
annihilating an electron in lead $j$ of wave vector $\bk$, spin $z$ component
$\sigma$, and energy $\epsilon_{\bk}$;
\begin{equation}
\label{H_dots}
H_{\text{dots}}=\sum_{i=1}^2 \bigl( \varepsilon_i n_i
  + U_i n_{i\uparrow} n_{i\downarrow} \bigr)
\end{equation}
describes the energetics of the dots in terms of their occupancies
$n_{i\sigma}=d_{i\sigma}^{\dag} d_{i\sigma}^{\pdag}$ and $n_i=n_{i\uparrow}
+n_{i\downarrow}$, where $d_{i\sigma}$ annihilates an electron of spin $z$
component $\sigma$ in the level of dot $i$ that lies closest to the common
Fermi energy of the two leads (taken to be $\varepsilon_F = 0$); and
\begin{equation}
\label{H_mix}
H_{\text{mix}}=\sum_{i,j,\bk,\sigma} V_{ij} \bigl( d_{i\sigma}^{\dag}
  c_{j\bk\sigma}+\text{H.c.} \bigr)
\end{equation}
accounts for electron tunneling between dots and leads. For simplicity, we take
real dot-lead couplings $V_{iL}=V_{iR}\equiv V_i/\sqrt{2}$, for which case the
dots interact only with one effective band formed by an even-parity combination
of $L$ and $R$ states. We assume a constant density of states $\rho=1/(2D)$
with half bandwidth $D$, so that the dot-lead tunneling is measured via the
hybridization widths $\Gamma_i=\pi\rho V_i^2$. At low bias, electron
transmission described by a Landauer-like formula\cite{Meir:92} gives a linear
conductance
\begin{equation}
\label{g}
g(T)=\frac{2e^2}{h}\int d\omega \biggr(\frac{-\partial f}{\partial\omega}
\biggr) \, \pi \sum_{i,j}\sqrt{\Gamma_i \Gamma_j} \, A_{ij}(\omega,T),
\end{equation}
where $f(\omega,T) = [\exp(\omega/T)+1]^{-1}$ is the Fermi-Dirac function and
$A_{ij}(\omega,T)=-\pi^{-1} \, \text{Im} \, G_{ij}(\omega,T)$ is the spectral
density corresponding to the retarded Green's function $G_{ij}(\omega,T) =
-i\int_0^{\infty}dt e^{i\omega t} \langle\{d_{i,\sigma}^{\pdag}(t), \,
d_{j,\sigma}^{\dag}(0)\}\rangle$.

We have studied this model using the numerical renormalization
group\cite{Bulla:08} with discretization parameter $\Lambda=2.5$, retaining at
least 1000 states after each iteration.\cite{Lane:08} This paper focuses on the
representative case of a strongly interacting dot 1 described by
$U_1=10\Gamma_1=0.5D$ and a dot-2 hybridization width $\Gamma_2=0.02D$.
We show the variation of physical properties with temperature $T$ and the
dot energies $\varepsilon_i$ (which should be experimentally tunable via
plunger gate voltages) for different values of $U_2$. We reiterate that in real
devices, $U_2$ will likely be fixed and $\Gamma_2$ instead will be varied by
raising or lowering tunnel barriers.

\begin{figure}[b]
\centerline{\includegraphics[width=3.3in]{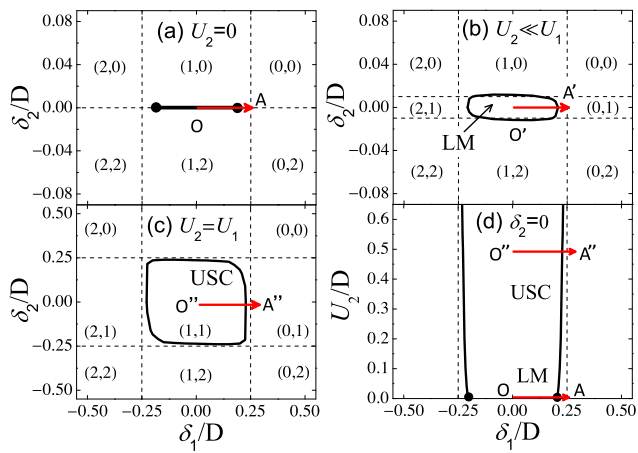}}
\caption{(Color online). Ground states of the isolated quantum dots
[dashed lines and dot occupancies $(\langle n_1\rangle, \langle n_2\rangle)$]
and phases of the full system for $\Gamma_1=0.05D$, $\Gamma_2=0.02D$ (solid
lines) vs level energies \mbox{$\delta_i=\varepsilon_i+\half U_i$}
measured from particle-hole symmetry for (a) $U_2=0$,
(b) \mbox{$U_2=0.02D\ll U_1$} and (c) $U_2=U_1$. (d) Phase diagram of the full
system on the $\delta_1$-$U_2$ plane at \mbox{$\delta_2=0$}, showing
local-moment (LM) and underscreened spin-1 Kondo (USC) regimes within the
free-moment phase. Filled circles in (a) and (d) indicate first-order QPTs
occurring only for $U_2=0$, while all other points along the phase boundaries
correspond to QPTs of the Kosterlitz-Thouless type. Arrows represent paths along
which data are plotted in Figs.\ \protect\ref{Fig2}--\protect\ref{Fig4} and
\protect\ref{Fig7}.}
\label{Fig2}
\end{figure}

It is instructive first to consider the dots isolated from the leads,
\textit{i.e.}, the limit $\Gamma_1=\Gamma_2=0$. Figures
\ref{Fig2}(a)--\ref{Fig2}(c) show $T=0$ occupancies
$(\langle n_1\rangle,\langle n_2\rangle)$ vs the level
energies $\delta_i=\varepsilon_i+\half U_i$ measured from particle-hole
symmetry for three values of the dot-2 Coulomb interaction strength:
$U_2=0$, $U_2=0.02D$ ($\ll U_1$) and $U_2=0.5D$ ($=U_1$). The value of
$\langle n_i\rangle$ jumps on crossing a dashed line representing
$\delta_i=\pm\half U_i$. For $U_2=0$ [Fig.\ \ref{Fig2}(a)],
$\langle n_2\rangle=1$ only when the dot-2 level lies precisely at the chemical
potential (along the line $\delta_2=0$) and the $\delta_1$-$\delta_2$ plane
divides into six two-dimensional regions. For $U_2>0$ [Figs.\ \ref{Fig2}(b,c)],
there are instead nine regions, including three in which dot 2 is singly
occupied and hence carries a magnetic moment.

When both dots are connected to the metallic leads ($\Gamma_1,\Gamma_2\not=0$),
the numerical renormalization-group solution reveals that most of the
$\delta_1$-$\delta_2$ plane is occupied by a \textit{strong-coupling phase} in
which all dot degrees of freedom are quenched at $T=0$. Within this phase, the
first-order QPTs present for isolated dots (dashed lines in Fig.\ \ref{Fig2})
are replaced by smooth crossovers between single-particle scattering of lead
electrons (wherever each dot is either empty or full, \textit{i.e.}, 
$|\delta_i|-\half U_i\gg\Gamma_i$ for $i=1$ \textit{and} 2) and many-body
Kondo physics (wherever one of the dots is singly occupied, \textit{i.e.},
$|\delta_i|-\half U_i\ll -\Gamma_i$ for $i = 1$ \textit{or} 2). However,
the region around the particle-hole-symmetric point $\delta_1=\delta_2=0$ forms
a distinct \textit{free-moment phase} in which a spin-\half\ degree of freedom
survives down to $T=0$. With increasing $U_2$, this free-moment phase
grows---primarily along the $\delta_2$ axis---as illustrated by the solid lines
in Fig.\ \ref{Fig2}.

The next two sections present physical properties along paths in parameter
space that are represented schematically by arrows in Fig.\ \ref{Fig2}. Each
path crosses the phase boundary at a location that can be parametrized as
$\varepsilon_1=\varepsilon_1^{\pm}(U_2,\varepsilon_2)$. (This notation
suppresses additional dependences of the phase boundaries on $U_1$ and on the
level widths $\Gamma_1$ and $\Gamma_2$, three quantities that are held constant
for all the results presented in this paper.) The Hamiltonian \eqref{H_full} is
invariant (up to a constant) under the particle-hole transformation
$c_{j\bk\sigma}^{\pdag} \to c_{j\bk\sigma}^{\dag}$,
$d_{i\sigma}^{\pdag} \to -d_{i\sigma}^{\dag}$,
$\epsilon_{j\bk} \to -\epsilon_{j\bk}$, and $\delta_i \to -\delta_i$.
This symmetry implies that the phase boundaries in Fig. \ref{Fig2} are invariant
under a simultaneous change in the sign of $\delta_1$ and $\delta_2$, or
equivalently that $\varepsilon_1^-(U_2,\varepsilon_2) =
-U_1 - \varepsilon_1^+(U_2,-U_2-\varepsilon_2)$.

\begin{figure}
\centerline{\includegraphics[width=3.45 in]{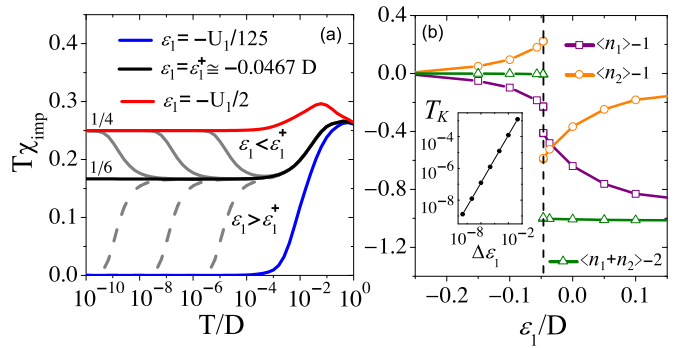}}
\caption{(Color online). Noninteracting dot 2, $U_2=\varepsilon_2=0$:
(a) $T\chi_{\text{imp}}$ vs $T$ for various values of $\varepsilon_1$ spanning
the QPT at $\varepsilon_1^+$. (b) $T=0$ dot occupancies relative to half
filling vs $\varepsilon_1$, with a vertical dashed line at
$\varepsilon_1=\varepsilon_1^+$. Inset: Evolution of the Kondo scale showing a
linear dependence on $\Delta\varepsilon_1=\varepsilon_1-\varepsilon_1^+$.}
\label{Fig3}
\end{figure}

\section{Zero versus weak dot-2 interactions}
\label{sec:zero_vs_weak_U2}

We begin by presenting the properties of the double-quantum-dot system when
Coulomb interactions in dot 2 are much weaker than in dot 1.  We will focus on
two specific cases, namely, $U_2=0$ and $U_2=\Gamma_2$. An understanding of
these cases will allow us to establish a connection with the large-$U_2$
regime in Sec.\ \ref{sec:small_vs_large_U2}.

\subsection{Noninteracting dot 2}

In the special case $U_2=0$, it is possible to integrate out the dot-2 degrees
of freedom, thereby mapping the double-dot setup to an effective one-impurity
Anderson model\cite{Dias:06} in which the interacting dot 1 hybridizes with a
conduction band described by a density of states
\begin{equation}
\rho_{\text{eff}}(\varepsilon) \simeq \frac{1}{2 D} \,
  \frac{(\varepsilon-\varepsilon_2)^2}{(\varepsilon-\varepsilon_2)^2+\Gamma_2^2}
\end{equation}
for $|\varepsilon|\ll D$. For $\varepsilon_2 \ne 0$, $\rho_{\text{eff}}(0)$ is
nonzero and the dot-1 degree of freedom is completely quenched at sufficiently
low temperatures. For $\varepsilon_2 = 0$, however,
$\rho_{\text{eff}}(\varepsilon)$ vanishes quadratically at $\varepsilon = 0$,
leading to a realization of the pseudogap Anderson model.\cite{Dias:06,Dias:08}
In the mapped problem, the free-moment phase can be interpreted as a region
of parameter space in which the loss of band states near the Fermi energy
prevents Kondo screening of the dot-1 spin.

This subsection reports results of calculations performed directly on the
double-dot model [Eq.\ \eqref{H_full}] with $U_2=0$. As found previously in
studies of the mapped problem,\cite{Dias:06,Dias:08} the free-moment phase is
restricted to $\varepsilon_2=0$, $\varepsilon_1^-<\varepsilon_1<\varepsilon_1^+$
[$\varepsilon_1^\pm (0,0)$ being denoted by filled circles in Figs.\
\ref{Fig2}(a,d)]. Figure \ref{Fig3}(a) shows the temperature variation of
$\chi_{\text{imp}}$, the contribution of the two dots (``impurities'') to the
magnetic susceptibility (defined and calculated in the usual
way\cite{Krishna-murthy:80}), for several values of $\varepsilon_1$
along path OA in Figs.\ \ref{Fig2}(a,d).
In the free-moment phase (\textit{e.g.}, $\varepsilon_1=-\half U_1$), a
doublet ground state survives down to $T=0$ with
$T\chi_{\text{imp}}=\frac{1}{4}$, characteristic of a free spin-\half.
In the strong-coupling phase (\textit{e.g.}, $\varepsilon_1=-U_1/125$), the
system instead has a singlet ground state and $\chi_{\text{imp}}$ (not just
$T\chi_{\text{imp}}$) vanishes as $T\to 0$. For $\varepsilon_1$ close to
$\varepsilon_1^+$, singlet and doublet ground states are quasi-degenerate and
$T\chi_{\text{imp}}\approx\frac{1}{6}$ within a window of temperatures above
some $T^*$; for $T\lesssim T^*$, there is a crossover to the low-temperature
behavior of one or other phase. The crossover scale $T^*$ vanishes continuously
on approach to the phase boundary from either side, and at $\varepsilon_1 =
\varepsilon_1^+$, $T\chi_{\text{imp}}=\frac{1}{6}$ down to $T=0$. The inset to
Fig.\ \ref{Fig3}(b) shows that the Kondo temperature $T_K$---proportional to
the crossover scale $T^*$ in the strong-coupling phase, and defined via the
standard condition\cite{Krishna-murthy:80}
$T_K \chi_{\text{imp}}(T_K)=0.0701$---vanishes linearly with
$\Delta\varepsilon_1=\varepsilon_1-\varepsilon_1^+$, as expected
at a first-order level-crossing QPT.

Further insight into the QPTs at $\varepsilon_1=\varepsilon_1^{\pm}(0,0)$ can
be gained by examining the dot occupancies $\langle n_i\rangle$ at zero
temperature. On approach to the QPT from either phase, the occupancies [Fig.\
\ref{Fig3}(b)] increasingly deviate from the values for isolated dots. Both
occupancies undergo a jump at $\varepsilon_1=\varepsilon_1^+$. The magnitude of 
the jump in $\langle n_1\rangle$ can be identified with the weight under a
delta-function peak in the dot-1 spectral density that passes through the Fermi
energy at the QPT.\cite{Dias:08} The limiting values of $\langle n_1\rangle$
and $\langle n_2\rangle$ on either side of the phase boundary, as well as the
magnitudes of the jumps at the QPT, are found to change significantly with
$U_1$, $\Gamma_1$, and $\Gamma_2$. However, the combined occupancy
$\langle n_1+n_2\rangle$ for $\varepsilon_2=0$ in all cases remains very close
to 2 throughout the free-moment phase, to 1 for all $\varepsilon_1 >
\varepsilon_1^+$, and to 3 for all $\varepsilon_1 < \varepsilon_1^-$.

In order to understand this striking behavior of $\langle n_1+n_2\rangle$, it
is useful to consider the wide-band limit in which $D$ greatly exceeds all
other energy scales. Here, $\langle n_1+n_2\rangle$ becomes identical to
$n_{\text{imp}}\equiv\langle N\rangle-\langle N\rangle_0$, where
$\langle N\rangle$ ($\langle N\rangle_0$) is the total number of electrons with
(without) the dots.\cite{Logan:09} One can find $n_{\text{imp}}$ using the
aforementioned mapping to a one-impurity pseudogap Anderson model, valid for
$U_2=\varepsilon_2=0$. In the free-moment phase of the pseudogap model,
particle-hole asymmetry is irrelevant\cite{Ingersent-pseudo} so
$n_{\text{imp}}(T=0)=2$; by contrast, particle-hole asymmetry is relevant in
the strong-coupling phase,\cite{Ingersent-pseudo} forcing
$n_{\text{imp}}(T=0)=1$ or $3$ depending on the sign of
$\delta_1\equiv\varepsilon_1+\half U_1$. These observations explain the
near-pinning of $\langle n_1+n_2\rangle$ away from the wide-band limit, where
$\langle n_1+n_2\rangle$ only approximately equals $n_{\text{imp}}$. They also
identify the differing response to particle-hole asymmetry in the two phases as
the underlying reason for the first-order nature of the $U_2=0$ QPTs.

\begin{figure}
\centerline{\includegraphics[width=3.45 in]{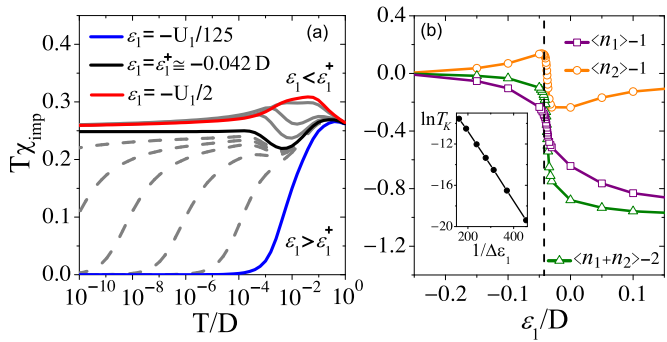}}
\caption{(Color online). Interacting dot 2, $U_2=-2\varepsilon_2=\Gamma_2$:
(a) $T\chi_{\text{imp}}$ vs $T$ for various values of $\varepsilon_1$ spanning
the QPT at $\varepsilon_1^+$. (b) $T=0$ dot occupancies relative to half
filling vs $\varepsilon_1$ with a vertical dashed line at
$\varepsilon_1=\varepsilon_1^+$. Inset: Evolution of the Kondo scale $T_K$,
showing $\ln T_K\propto 1/\Delta\varepsilon_1$ where
$\Delta \varepsilon_1=\varepsilon_1-\varepsilon_1^+$.}
\label{Fig4}
\end{figure}

\subsection{Weakly interacting dot 2}
\label{subsec:small_U2}

Now we turn to the case $U_2=\Gamma_2$ representative of the crossover 
from a resonant dot 2 to an interacting one. The mapping to an effective
one-impurity model breaks down for $U_2 \ne 0$, so the full double-dot
model must be solved directly.

Figure \ref{Fig4}(a) plots $T\chi_{\text{imp}}$ vs $T$ at different points
along path O$^{\prime}$A$^{\prime}$ in Fig.\ \ref{Fig2}(b). Deep in the
strong-coupling phase (\textit{e.g.}, $\varepsilon_1=-U_1/125$) the system
passes with decreasing temperature directly from a local-moment regime
($T\chi_{\text{imp}}=\frac{1}{4}$) to the strong-coupling limit
(T$\chi_{\text{imp}}=0$); just as for $U_2=0$, $\chi_{\text{imp}}(T=0)=0$.
For $\varepsilon_1$ just above $\varepsilon_1^+$ [\textit{e.g.}, uppermost
dashed line in Fig.\ \ref{Fig4}(a)], $T\chi_{\text{imp}}$ instead evolves with
decreasing $T$ from near $\frac{1}{4}$ towards the value $\frac{1}{6}$
characterizing the $U_2=0$ QPT (a tendency seen more clearly\cite{Lane:08} for
$0<U_2\ll\Gamma_2$), then rises and reaches a plateau near $\frac{1}{4}$ before
finally decreasing to zero. The manner in which $T\chi_{\text{imp}}\rightarrow0$
as $T\rightarrow0$ is identical to that in the Kondo regime of the conventional
Anderson model,\cite{Krishna-murthy:80} with
$\chi_{\text{imp}}(T=0) \simeq 0.1/T_K$ and $T_K$ varying exponentially with
$1/(\varepsilon_1-\varepsilon_1^+)$ [inset to Fig.\ \ref{Fig4}(b)]. For
$\varepsilon_1<\varepsilon_1^+$, $T\chi_{\text{imp}}$ approaches the free-moment
value $\frac{1}{4}$ from above, but there is no temperature scale that vanishes
on approach to the phase boundary. These behaviors are all indicative of the
Kosterlitz-Thouless nature of the QPT, which holds for any $U_2>0$ (with the
sole exception of the first-order QPTs that arise from parity conservation in
the special case of two identical Kondo-regime dots\cite{Logan:09}).
Like the ferromagnetic Kondo model, whose properties it closely parallels, the
small-$U_2$ free-moment phase exhibits singular Fermi liquid behavior with a
quasiparticle density of states that diverges at the Fermi
energy.\cite{Mehta:05,phase-shifts}

The dot occupancies for $U_2=\Gamma_2$ [Fig.\ \ref{Fig4}(b)] show generally
the same trends vs $\varepsilon_1$ as found for $U_2=0$ [Fig. \ref{Fig3}(b)],
with the significant difference that there are no jumps. Since
particle-hole asymmetry is a marginal perturbation in the conventional Anderson
model,\cite{Krishna-murthy:80} $n_{\text{imp}}(T=0)$ varies continuously with
$\varepsilon_1$, and there is no pinning of $\langle n_1+n_2\rangle$ in either
phase.

Comparison between Figs.\ \ref{Fig3} and \ref{Fig4} shows that for
$U_2\lesssim\Gamma_2$, the properties retain their $U_2=0$ pseudogap character
provided that the system is sufficiently far from the location $T=0$,
$\varepsilon_1 = \varepsilon_1^{\pm}$ of the QPT. With decreasing $U_2$ (not
shown), the pseudogap behavior progressively extends to lower temperatures
and/or smaller $|\varepsilon_1-\varepsilon_1^{\pm}|$.

\begin{figure}[b]
\centerline{\includegraphics[width=3.5 in]{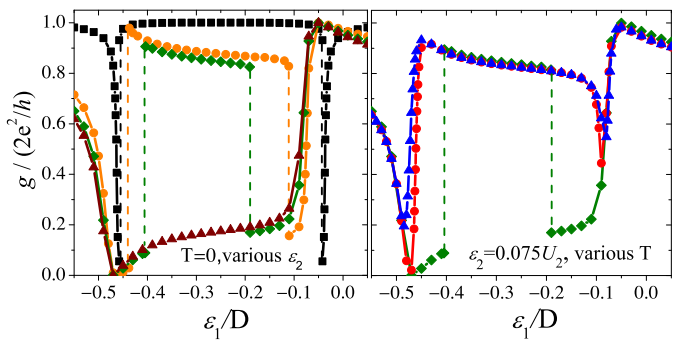}}
\caption{(Color online). Linear conductance $g$ vs $\varepsilon_1$ for
$U_2=\Gamma_2$. Left: Temperature $T=0$ and scaled dot-2 level energies
$\varepsilon_2/U_2 = -0.5$ ($\blacksquare$), 0 ({\Large $\bullet$}), 0.075
($\blacklozenge$), and 0.11 ($\blacktriangle$). Right: $\varepsilon_2=0.075 U_2$
and scaled temperatures $T/T_{K0}=0$ ($\blacklozenge$), 0.0057
({\Large $\bullet$}), and 0.228 ($\blacktriangle$).}
\label{Fig5}
\end{figure}

The physical property most likely to be accessible in experiments is
the electrical conductance between the left and right leads. Figure
\ref{Fig5}(a) shows the linear conductance $g$ [Eq.\ \eqref{g}] as a function
of $\varepsilon_1$ for $U_2=\Gamma_2$, $T=0$, and four values of
$\varepsilon_2$. Deep in the free-moment phase (around
$\varepsilon_i=-\half U_i$), dot 1 is in Coulomb blockade and since there
is no Kondo effect and hence no Kondo resonance, transport takes place solely
through dot 2. For fixed $\varepsilon_1$ near $-\half U_1=-0.25D$, the
zero-temperature conductance decreases from its unitary limit $g=2e^2/h$ as
$\varepsilon_2$ is varied from $-\half U_2$ (squares) to higher (circles
and diamonds) or lower values, while for fixed $\varepsilon_2$ near
$-\half U_2$, the system passes through a QPT at
$\varepsilon_1=\varepsilon_1^{\pm}$, where $g$ undergoes a jump. For
$\varepsilon_1$ right above $\varepsilon_1^+$ or right below $\varepsilon_1^-$,
there is a Kondo effect centered primarily on dot 1, and interference between
transport through the two dots causes $g$ to decrease abruptly. On moving
deeper into the strong-coupling phase, the dot-1 occupancy moves further from
unity, interference from transport through dot 1 is reduced, and $g$ rises
again. The preceding picture holds until dot 2 becomes sufficiently
particle-hole asymmetric that the strong-coupling phase spans all values of
$\varepsilon_1$, and $g$ vs $\varepsilon_1$ shows no sign of any QPT
[triangles in Fig.\ \ref{Fig5}(a)].

The conductance signatures of the QPT persist to $T>0$, as illustrated in
Fig.\ \ref{Fig5}(b), which plots $g$ vs $\varepsilon_1$ for $U_2=\Gamma_2$,
$\varepsilon_2=0.075 U_2$, and three temperatures specified in the caption as
multiples of $T_{K0}=7\times10^{-4}D$: the Kondo scale when dot 2 is isolated
($\Gamma_2=0$) and dot 1 is at particle-hole symmetry
($\varepsilon_1=-\half U_1)$. The foremost effect of increasing $T$ is a
progressive suppression of the Kondo effect, leading to a smoothing and
weakening of the conductance dips in the vicinity of the QPTs, as well as
shifts in positions of the local minima in $g$ to larger values of
$|\varepsilon_1+\half U_1|$.

\section{Weak versus strong dot-2 interactions}
\label{sec:small_vs_large_U2}

In this section, we compare the regime $U_2\lesssim\Gamma_2$ described above
with the one $U_2\gg\Gamma_2$ studied in most previous work on Kondo physics
in parallel double quantum dots. We show that these regimes have very different
spin correlations between the different components of the double-quantum-dot
device. Furthermore, the regimes can be distinguished experimentally through
linear conductance measurements. 

\subsection{Spin correlations}

Insight into the connection between the regimes of small and large
$U_2/\Gamma_2$ can be gained from the static spin-spin correlation
$\langle \mathbf{S}_i \cdot \mathbf{S}_{\text{leads}} \rangle$ between dot $i$
and the leads, as well as from the interdot correlation
$\langle \mathbf{S}_1 \cdot \mathbf{S}_{2} \rangle$. Here, 
$\mathbf{S}_i=\half\sum_{\sigma,\sigma'} d_{i\sigma}^{\dag}
\boldsymbol{\sigma}_{\sigma,\sigma'} d_{i\sigma'}^{\pdag}$ and
$\mathbf{S}_{\text{leads}}=\half\sum_{j,\bk,\bk',\sigma,\sigma'}
c_{j\bk\sigma}^{\dag} \boldsymbol{\sigma}_{\sigma,\sigma'}
c_{j\bk'\sigma'}^{\pdag}$, where $\boldsymbol{\sigma}$ is a vector of
Pauli matrices.

\begin{figure}
\centerline{\includegraphics[width=2 in]{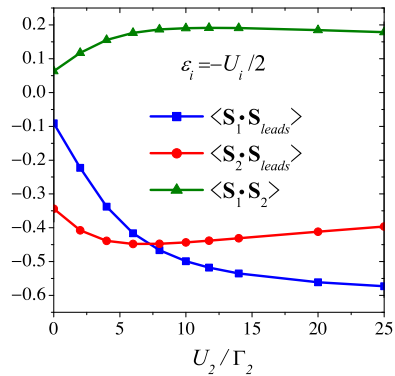}}

\caption{(Color online). Dot-lead and dot-dot spin correlations vs scaled
interaction strength $U_2/\Gamma_2$, determined at zero temperature for level
energies $\varepsilon_i =- \half U_i$, \textit{i.e.}, at the center of the
free-moment phase. Increasing $U_2$ from zero enhances the entanglement between
dot 1 and the other parts of the system as the residual spin-\half\ degree of
freedom evolves from being localized on dot 1 (for $U_2=0$) to being
distributed throughout the system (for $U_2\gg\Gamma_2)$.}
\label{Fig6}
\end{figure}

Figure \ref{Fig6} shows the $T=0$ spin-spin correlations vs $U_2/\Gamma_2$ for
fixed $\Gamma_2=0.02D$ with both dots at particle-hole symmetry, \textit{i.e.},
at the center of the free-moment phase.\cite{spin-correlations} For
$U_2=\varepsilon_2=0$, spin-0 and spin-\half\ configurations of dot 2 should be
equally probable, whereas dot 1 is expected to have a well-defined spin-\half\
at low temperatures. The facts that
$\langle \mathbf{S}_1 \cdot \mathbf{S}_{\text{leads}}\rangle$ is much smaller
in magnitude than $\langle \mathbf{S}_2 \cdot \mathbf{S}_{\text{leads}}\rangle$,
and that the latter quantity is close to the value $\half \times (-\frac{3}{4})
= -\frac{3}{8}$ it would take if dot 1 were absent from the system, indicate
that for $U_2=0$ the residual spin-\half\ degree of freedom is located
primarily on dot 1, which is almost decoupled from other parts of the system.

Increasing $U_2$ enhances the magnetic character of dot 2 and so strengthens
both the dot's antiferromagnetic correlation with the leads and (via an
effective RKKY interaction\cite{Zitko:06,Logan:09}) its ferromagnetic
correlation with dot 1. There is an even more pronounced growth in the
antiferromagnetic correlation between dot 1 and the leads. These trends
continue until $U_2/\Gamma_2$ becomes of order 5, by which point each dot
carries a well-defined spin-\half. To good approximation, these spins combine
to form a triplet that is partially Kondo-screened by the leads, to yield a
strongly entangled spin-\half\ ground state.\cite{Logan:09,USC} Since the
effective exchange interaction between dot 2 and the leads is
proportional\cite{Zitko:06} to $1/U_2$, further increase of
$U_2/\Gamma_2$ beyond about 5 results in a gradual reduction in the magnitudes
of both $\langle \mathbf{S}_1 \cdot \mathbf{S}_{\text{leads}}\rangle$ and
$\langle \mathbf{S}_1 \cdot \mathbf{S}_2\rangle$.                   

\subsection{Transport properties}

Although the regimes $U_2\ll\Gamma_2$ and $U_2\gg\Gamma_2$ feature very
different spin correlations, they belong to the same phase and therefore have
qualitatively the same asymptotic low-temperature properties.\cite{phase-shifts}
The question remains whether the two regimes may be distinguished through their
behavior at higher $T$.

\begin{figure}
\centerline{\includegraphics[width=3.4 in]{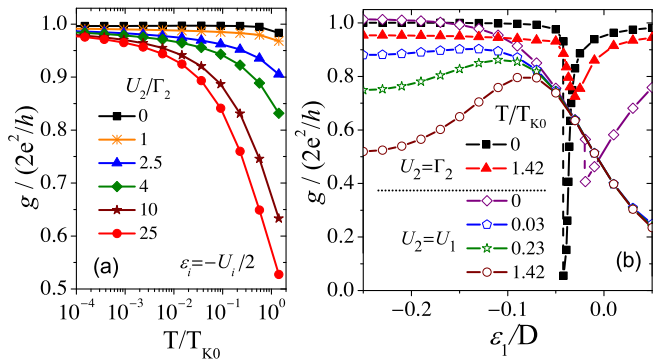}}
\caption{(Color online). Conductance $g$  plotted (a) vs $T$ at particle-hole
symmetry ($\varepsilon_i=-\half U_i$); and (b) vs $\varepsilon_1$ for
$\varepsilon_2=-\half U_2$, comparing the local-moment and underscreened
spin-1 Kondo regimes of the free-moment phase, represented by $U_2=\Gamma_2$ and
$U_2=U_1$, respectively.}
\label{Fig7}
\end{figure}

Figure \ref{Fig7}(a) shows $g$ vs $T$ at the particle-hole-symmetric point
$\varepsilon_i=-\half U_i$ for six values of $U_2$. For
$U_2\gg\Gamma_2$, the conductance drops significantly below its unitary limit
once the temperature rises above the characteristic scale $T_K^{S=1}$ of the
spin-1 Kondo effect, which is\cite{Logan:09} of order $T_{K0}$. 
For $U_2\lesssim\Gamma_2$, there is no Kondo physics in the free-moment phase
and $g$ remains close to $2e^2/h$ up to much higher temperatures of order
$\Gamma_2$.

Figure \ref{Fig7}(b) plots $g$ vs $\varepsilon_1$ at different temperatures
for $U_2=-2\varepsilon_2=\Gamma_2$ (path O$^{\prime}$A$^{\prime}$ in
Fig.\ \ref{Fig2}) and for $U_2=-2\varepsilon_2=U_1$ (path
O$^{\prime\prime}$A$^{\prime\prime}$). Just as in Fig.\ \ref{Fig7}(a),
the $T$ dependence of the conductance in the free-moment phase is much weaker
for $U_2\lesssim\Gamma_2$ than for $U_2\gg\Gamma_2$. Near particle-hole
symmetry ($\varepsilon_1=-0.25D$), the latter regime has
$d^2g/d\varepsilon_1^2>0$ at all but the very lowest temperatures, reflecting
the $\varepsilon_1$ dependence\cite{Logan:09} of $T_K^{S=1}$, whereas
$d^2g/d\varepsilon_1^2\le 0$ in the local-moment case.

Similar trends to those shown in Fig.\ \ref{Fig7} are found for other
choices of $\varepsilon_1$ and $\varepsilon_2$ that place the system in the
free-moment phase. We conclude that the local-moment and underscreened spin-1
Kondo regimes can be clearly differentiated via their conductance at
temperatures (of order the typical Kondo scale $T_{K0}$) that should be readily
attainable in experiments. 

\section{Summary}
\label{sec:summary}

We have studied two quantum dots coupled in parallel to metallic leads,
focusing on situations where ``dot 2'' has a weaker on-site Coulomb interaction
than ``dot 1'': $U_2<U_1$. For $U_2\lesssim\Gamma_2$, the tunneling width of
the dot-2 level, the properties still reflect the pseudogap Kondo physics found
previously for $U_2=0$. For all $U_2>0$, Kondo-screened and free-moment phases
are separated by quantum phase transitions of the Kosterlitz-Thouless type that
have signatures in the electrical conductance up to experimentally accessible
temperatures.

In the free-moment phase, conductance measurements can also distinguish the
small-$U_2$ regime, in which dot 1 carries a spin-$\half$ and is essentially 
disconnected from the rest of the system, from the regime $U_2\gg\Gamma_2$ in
which both dots contain strong electron correlations and their combined spin
is partially screened by the leads. Given the feasibility of tuning between
these two cases---and of crossing into the Kondo phase (above an underlying
zero-temperature transition)---by adjusting just one gate voltage on each dot,
this system offers fascinating possibilities for controlled experimental study
of quantum phase transitions and of variations in the strength and spatial
distribution of entanglement in singular Fermi liquids.

\begin{acknowledgments}
We thank D.\ Logan for helpful discussions. This work was supported by NSF
DMR Grants 0710540, 0710581, 1107814, and 1108285 in the United States, and
by CNPq and by FAPESP Grant 2010/20804-9 in Brazil.
\end{acknowledgments}


\begin{thebibliography}{99}
\bibitem{phase-shifts}
Analysis of the scattering phase shifts indicates that the free-moment phase
exhibits singular Fermi liquid behavior for all $U_2>0$.
\bibitem{spin-correlations}
Within the numerical renormalization-group (NRG) approach, the spin of the
leads is calculated as $\mathbf{S}_{\text{leads}}=\half\sum_{n=0}^N
\sum_{\sigma,\sigma'} f_{n\sigma}^{\dag} \boldsymbol{\sigma}_{\sigma,\sigma'}
f_{n\sigma'}^{\pdag}$ where $f_{n\sigma}$ annihilates an electron of spin $z$
component $\sigma$ on site $n$ of the linear chain
resulting\protect\cite{Bulla:08} from the logarithmic discretization of
$H_{\mathrm{leads}}$ [Eq.\ \protect\eqref{H_leads}]. The correlations plotted
in Fig.\ \protect\ref{Fig6} are limiting values for high NRG iteration numbers
$N$ corresponding to temperatures $T\to0$.
\end{thebibliography}
\end{document}